\documentclass[twocolumn,prl,tightenlines,superscriptaddress,showpacs,nofootinbib]{revtex4-1}
\usepackage{amsmath}
\usepackage{amssymb,amsfonts,latexsym}
\usepackage{bm}
\usepackage[mathcal]{euscript}
\usepackage{graphicx}
\usepackage{epsfig}
\usepackage{color}
\usepackage{xfrac}
\usepackage{comment}
\usepackage{pdfpages}
\usepackage{pgffor}
\makeatletter
\AtBeginDocument{\let\LS@rot\@undefined}
\makeatother
\def\supplementfilename{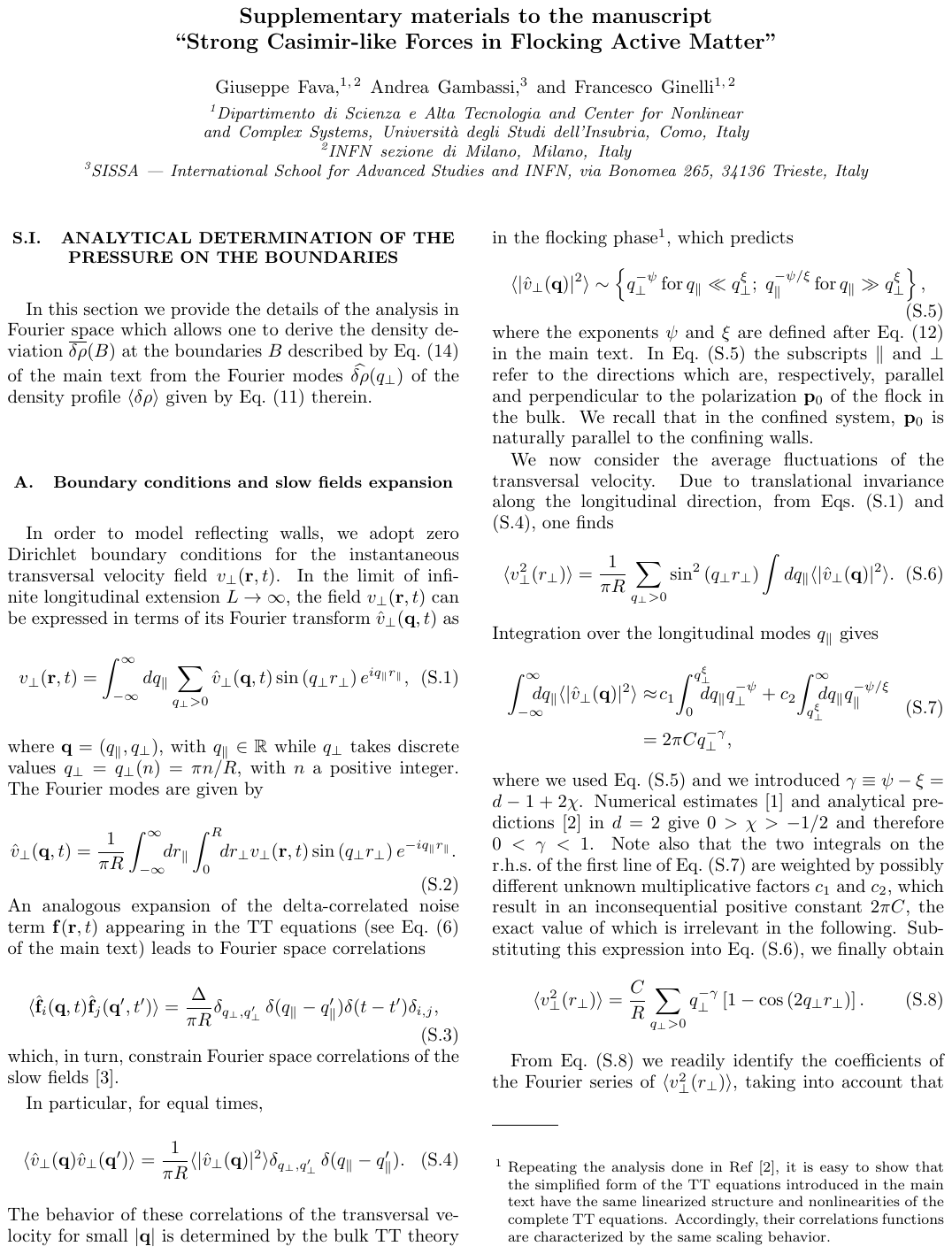}
\pdfximage{\supplementfilename}
\def\numbersupplementpages{\the\pdflastximagepages}
\newif\ifarXiv
\arXivtrue

\begin{document}

\title{Strong Casimir-like Forces in Flocking Active Matter}

\author{Giuseppe Fava}
\affiliation{Dipartimento di Scienza e Alta Tecnologia and Center for Nonlinear and Complex Systems, Universit{\`a} degli Studi dell'Insubria, Como, Italy}
\affiliation{INFN sezione di Milano, Milano, Italy}

\author{Andrea Gambassi}
\affiliation{SISSA --- International School for Advanced Studies and INFN, via Bonomea 265, 34136 Trieste, Italy}

\author{Francesco Ginelli}
\affiliation{Dipartimento di Scienza e Alta Tecnologia and Center for Nonlinear and Complex Systems, Universit{\`a} degli Studi dell'Insubria, Como, Italy}
\affiliation{INFN sezione di Milano, Milano, Italy}

\date{\today}

\begin{abstract}
Confining in space the equilibrium fluctuations of statistical systems
with long-range correlations is known to result into effective forces
on the boundaries. Here we demonstrate the occurrence of 
Casimir-like forces in the non-equilibrium context provided by
flocking active matter. In particular, we consider a system of
aligning self-propelled particles in two spatial dimensions, which are
transversally confined by reflecting or partially reflecting walls. We
show that in the ordered flocking phase this confined active {\it vectorial} fluid is characterized by 
{\it extensive} boundary layers, as
opposed to the finite ones usually observed in confined {\it scalar}
active matter. Moreover,  
a finite-size, fluctuation-induced contribution to the pressure on the wall emerges, 
which decays slowly and algebraically upon increasing the distance between
the walls. We explain our
findings -- which display a certain degree of universality -- within a hydrodynamic 
description of the density and velocity fields.
\end{abstract}

\maketitle


\emph{Introduction.---}Flocking, i.e., the collective motion exhibited by certain active matter
capable of spontaneously breaking its
rotational symmetry,
is a ubiquitous phenomenon,
which is observed in a wide variety 
of living systems and 
across various scales \cite{Ramaswamy}. Examples range from animal
groups \cite{Fish,Cavagna2010} to bacterial colonies \cite{bacteria1,bacteria2, bacteria3} and cellular migrations \cite{Giavazzi}, down to the cooperative behavior of molecular motors and biopolymers at the subcellular level \cite{Shaller}.

The bulk flocking state,
described by the celebrated Toner-Tu theory (TT) \cite{TT1,TT2}, is characterized by a strongly fluctuating ordered
phase with
long-range correlations in its slow fields, i.e., density and
orientation \cite{Ginelli2016}. 
While our understanding of this bulk behaviour is now
fairly complete, at least when the surrounding fluid may be safely
neglected (the so-called ``dry'' approximation \cite{Dry}), very little
is known regarding the collective behaviour of flocking
active matter subject to spatial confinement. 

This problem is 
relevant for many experimental
realizations with active colloids, where confinement by hard boundaries (e.g., 
within a ring \cite{Bartolo}) 
is practically unavoidable. 
Moreover,
confinement along one spatial direction has also been employed in
numerical investigations of collective motion models \cite{TT3, Benoit2019} as a way to control the diffusion of the mean flock orientation in finite systems
without imposing a global symmetry breaking field \cite{Nikos}.

Confinement effects have been so far mostly investigated in scalar active
matter (where the density field is the unique hydrodynamic mode),
either in ``dry'' systems \cite{Kudrolli2008, Deseigne2010, Ray, Marchetti2014a,
  Marchetti2014b, Fily2014, ABPs, Caprini2018, Das2018} 
or active suspensions \cite{Goldstein2012, Goldstein2013, Boffetta2022}. 
Interestingly,
recent work in active nematics \cite{ConfinedAN} and confined active
Brownian particles \cite{Tailleur2022a, Tailleur2022b}
revealed non-trivial long-range effects on  
the bulk dynamics of the system. 

In this study, we focus on confined flocking active matter and
investigate the onset of Casimir-like forces. 
These are long-range forces arising from the confinement of
a fluctuating correlated field \cite{KB-94,Gambassi}. 
They may be of {\it quantum} origin, as
noted by Casimir \cite{Casimir}, who considered the quantum vacuum fluctuations of the electromagnetic
field confined between two parallel conducting plates, but
also of thermal origin. This is the case, e.g., when one confines a classical fluctuating field
characterized by a diverging correlation length, such as in a binary liquid mixture approaching its demixing critical point \cite{deGennes}.
In both cases, the (free) energy of the confined medium
depends on the separation $R$ between the 
boundaries (as the set of allowed eigenmodes of the fluctuating fields
changes) and 
a weak, but measurable force emerges on the
boundary walls, which may decay algebraically 
upon increasing 
$R$. 
Casimir-like forces of thermal origin may also be found, inter alia, far from a
critical point, in equilibrium systems such as liquid crystals
\cite{PelitiProst}, where long-ranged soft modes emerge from the
spontaneous breaking of a continuous symmetry.

Statistical systems far from equilibrium
may also develop 
spatially correlated fluctuations, in some cases of non-thermal nature.
These fluctuations, when spatially confined by boundaries or inclusions, may give rise to short-range effective forces \cite{AngelPRL11}, possibly long-range Casimir-like forces \cite{AKK-15,CBM-06,BettoloMarconi,PRS-14}, and generalizations thereof \cite{NG-04}.
For instance, Ref.~\cite{BettoloMarconi} suggests that
long-range forces may arise upon confining a reaction-diffusion
systems near an absorbing phase transition. 
Casimir-like forces have been also investigated numerically in scalar active matter such as
run-and-tumble particles \cite{Ray}, but these systems lack
genuine long-range correlations and the forces arise from geometrical
constraints rather than from a collective behavior.
Here, instead,  
we predict truly collective long-ranged forces which 
arise upon confining flocking active
matter between flat 
and parallel
reflecting boundaries -- either elastic or
partially inelastic -- along one
direction transversal to that of
collective motion. 
Numerical simulations, supported by analytical arguments, show that non-equilibrium fluctuations induce unusually strong
Casimir-like forces, characterized by a remarkably 
slow algebraic decay. 

\emph{Model and numerical results.---}We consider a flocking system of $N$ particles in $d = 2$ spatial dimensions
confined between two parallel 
hard walls 
at distance $R$. We assume reflecting boundary conditions at the walls,
so that the symmetry-broken mean velocity of the flock fluctuates around the
direction parallel to them, which we refer to as longitudinal ($\|$). 
Periodic boundary conditions are
assumed
in that
direction, of
(large) length $L$.

The overdamped, time-discrete particle 
dynamics is provided 
by the prototypical Vicsek model (VM) \cite{vicsek1995}. 
The self-propelled particles are described by their position ${\bf r}_i^t$ and orientation 
${\bf n}_i^t\!=\!(\cos \theta_i^t, \sin  \theta_i^t)$ at time $t$, with  $i\!=\!1,\ldots, N$. 
They 
tend to align with their
local neighbours while moving with constant velocity $v_{\rm 0}{\bf
  n}_i^t$, i.e.,
\begin{equation}
{\theta}_i^{t+1}=\textrm{arg}\left(\sum_{j \in S_i} {\bf n}_j^t
                    \right) +\eta \xi_i^t, 
\quad\quad 
{\bf r}_i^{t+1}={\bf r}_i^t + v_{\rm 0} {\bf n}_i^{t+1}, 
\label{VM}
\end{equation}
where $\textrm{arg}({\bf v})$ is
the angle defined by the vector 
${\bf v}$, $S_i$ is the unit circle centered around the particle $i$ and
$\xi_i^t$ is a zero-average, delta-correlated noise uniformly drawn
from the interval $[-\pi, \pi]$. We fix the speed $v_0\!=\!0.5$ so that
we are left with two control
parameters, i.e., the noise amplitude $\eta$ and the total particle density
$\rho_0\!=\!N/(LR)$. 

The reflecting boundaries 
reverse the
perpendicular component (denoted by the
subscript $\perp$) of the particle orientation w.r.t.~the walls. 
Algorithmically, whenever the
dynamics (\ref{VM}) renders a particle position, with transversal component $r_\perp$,
outside the 
region $r_\perp \!\in\! [0,R]$, 
we apply the collision rule $n_\perp \!\!\to\! -\beta n_\perp$ and $r_\perp
\!\!\to\! 2 B \!- \beta r_\perp$ with either $B\!=\!0$ (left boundary) or
$B\!=\!R$ (right boundary), with $0<\beta \leq 1$ a collision parameter
modelling the tendency of particles to move away from the wall after a
collision. Since the VM describes the overdamped
dynamics of self-propelled particles which locally
inject and dissipate energy, the unit norm of the orientation
after a collision is
instantaneously
restored by the corresponding increase of its longitudinal component
\cite{note-npara}.

The mechanical pressure is defined as the force per unit length exerted by the
confined active fluid on the walls. For generic active fluids it is
not a state function and its exact 
form depends on the
details of the interaction of the active particles with the wall
\cite{TailleurP, TailleurP2}. In particular, it depends on the mechanisms
involved in the reorientation of the transverse self-propulsion
direction, which may also involve momentum transfer to the surrounding viscous fluid and/or substrate. These non-universal features are not
captured by the overdamped dynamics of the Vicsek model.
However, assuming that a local mechanical interaction prevents 
the active particle from crossing the boundaries, 
the mechanical pressure can be expressed as minus the force
per unit length that the walls exert 
on the particles positional degree of freedom. 

Consider, for instance, a wall which confines the particles in the region with $r_\perp < R$. 
At the mesoscopic scale,
assuming traslational invariance along the longitudinal direction and 
a sharp potential $U(r_\perp)$ in the transversal direction such that $U(r_\perp
<R) =0$ and $\partial_\perp U(r_\perp)   \approx \delta(r_\bot - R)$,
one obtains the following average pressure \cite{TailleurP} in terms of the local particle density $\rho({\bf r},t)$:
\begin{equation}
\label{eq:pressure}
\overline{P}_b(R) = \left\langle \int_{\bar{x}}^\infty \partial_\perp
  U(r_\perp) \rho(r_\perp) dr_\perp \right\rangle \approx  \langle  \rho(R) \rangle,
\end{equation}
where $\bar{x}$ is an arbitrary bulk point in $(0,R)$ and $\langle \cdot
\rangle$ denotes time (or, due to ergodicity, ensemble)
average. 

Microscopically, this amounts at
defining the pressure as the average number of
collisions per unit length, $\overline{P}_b=\langle
  M(t)\rangle/L$, with $M(t)$ the instantaneous number of collisions
  on the confining wall. In the following, we will adopt this definition, but we
  have also verified that an alternative definition in terms of the
  nominal momentum exchanged in the collisions yields the same results
  (up to a constant scaling factor) \cite{SM}. 

\begin{figure}[t!]
\includegraphics[width=0.48\textwidth]{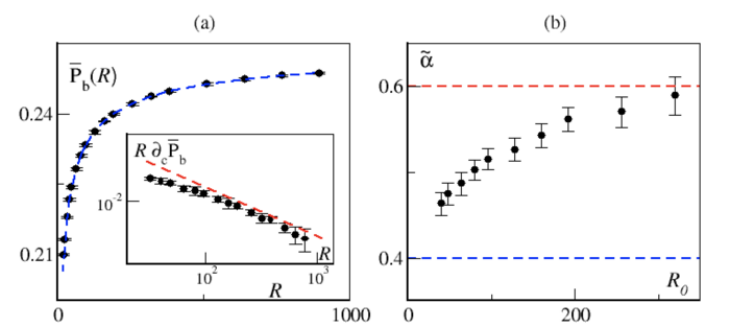}
\caption{(a) Time-averaged mechanical pressure exerted by the polar
  fluid on the
  confining walls as a function of their separation
  $R$ for $\eta\!=\!0.21$. The dashed blue line is the best fit by
  Eq.~(\ref{eq:pow}). Inset: Rescaled centered finite difference of $\overline{P}_b$
  as a function of $R$. The dashed red line indicates the best power-law
  fit for
  $R\!>\!R_0=80$ (see text). (b) Estimates
  of the effective exponent $\tilde{\alpha}(R_0)$ when the least squares fit is
  restricted in the region $R>R_0$. The dashed horizontal lines mark
  the interval $[0.4,0.6]$.
}
\label{fig1}
\end{figure}

The detailed phase diagram of the VM has been worked out in 
Refs.~\cite{Chate1,Chate2,VicsekPD}: it displays a homogeneous ordered phase and
a disordered one, separated by a coexistence region (see also Ref. \cite{Ihle2}).
Here we consider $\rho_0 \!= \!2.0$ and noise amplitude $\eta \! = \!0.21$, corresponding to the bulk homogeneous
ordered phase. Fixing the longitudinal size
$L\!=\!512$, we have measured
the average mechanical pressure $\overline{P}_b$ in systems of various transverse
sizes $R$. We first consider perfectly reflecting walls, i.e., $\beta =1$. Our results, reported in Fig.~\ref{fig1}(a), show that the
pressure increases upon increasing 
$R$ and that it approaches
the  
bulk value $\overline{P}_b(\infty)$ as $R\!\to\!\infty$. This value
characterizes the mechanical pressure exerted on a reflecting boundary by a
semi-infinite system. In equilibrium, 
it typically originates from the bulk contribution to the free energy. 

Our data, presented in Fig.~\ref{fig1}(a), are well fitted by the
  three-parameter function
\begin{equation}
\overline{P}_b(R) =\overline{P}_b(\infty) - C R^{-\alpha},
\label{eq:pow}
\end{equation}
with $\overline{P}_b(\infty)$ and $C$ positive and $\alpha \approx 0.5$.
By varying the
density $\rho_0$ and the noise amplitude
$\eta$, we also verified that this result holds generically 
in the flocking phase, with the boundary pressure increasing monotonically with either $\rho_0$ and $\eta$ \cite{SM}.

In the typical Casimir setup in which 
two infinite reflecting walls, separated by a distance $R$, are immersed in a much larger
flocking active fluid of transverse size 
$L_\perp \!\gg\! R$, the bulk contributions
exerted on the two faces of the walls cancel out and one is left with
the net pressure 
\begin{equation}
\Delta \overline{P}_b(R) \equiv
\overline{P}_b(\infty)\!-\!\overline{P}_b(R) = C R^{-\alpha},
\label{eq:Casimir}
\end{equation}
due to the interaction between the confining walls.
This shows that two infinite and
parallel reflecting walls, separated by a distance $R$ and immersed in
an active polar fluid in the
flocking phase experience an \emph{attractive}, long-range
Casimir-like pressure.

The {\it Casimir
  exponent} $\alpha$ can be estimated by considering centered finite difference 
$\partial_c f(x)$ of the average pressure,
which approximates the first derivative of $f(x)$ up to corrections of order
$f'''(x)$. Assuming that Eq.~(\ref{eq:pow}) holds, we have
$ R \partial_c \overline{P}_b(R) \approx\alpha C R^{-\alpha}$,
which renders the estimate $\alpha \!=\! 0.51(2)$ by fitting the data with $R \ge R_0 = 80$ (see inset of Fig.~\ref{fig1}(a)). A closer inspection, however, reveals a
systematic dependence of the estimated exponent $\tilde{\alpha}(R_0)$ on the smallest value $R_0$ of 
$R$ actually included while fitting the data, thus excluding the smaller
transversal separations $R<R_0$ from the least squares regression algorithm. 
This behavior, shown in Fig.~\ref{fig1}(b), suggests a finite-size crossover within
the range $0.4 \lesssim \tilde{\alpha}(R_0)\lesssim 0.6$. 

We finally consider partially reflecting boundaries. 
Simulations with $\beta\!=\!0.5$ and $0.25$ 
(shown in~\cite{SM})
exhibit the same long-range behavior of the mechanical pressure,
confirming a certain degree of universality, at least for repelling
confinements, with particles moving away from the wall after the collision.

%
%
\begin{figure}[t!]
\includegraphics[width=0.48\textwidth]{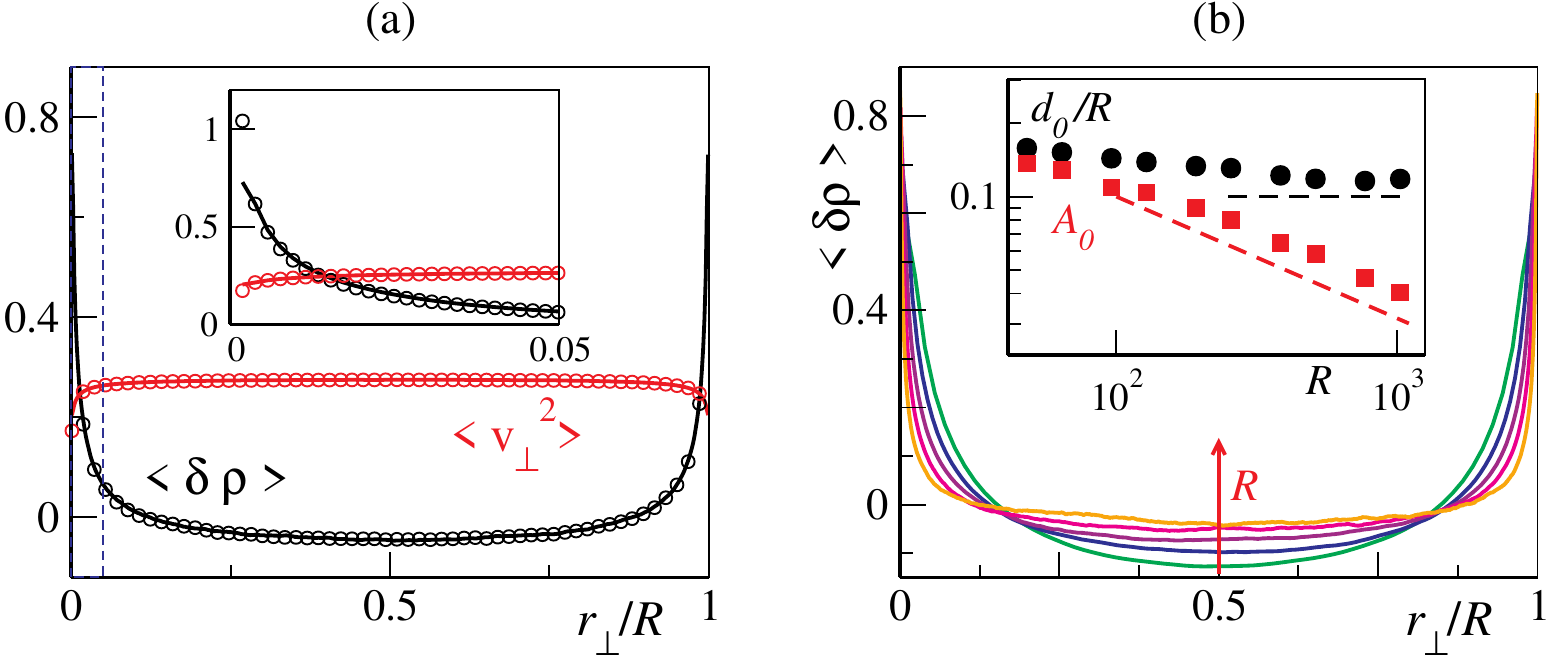}
\caption{(a) Average density deviations (black) and transversal
  velocity fluctuations (red) for $R\!=\!512$ as a function of the rescaled
  transverse coordinate $r_\perp/R$.
The case of elastic (solid lines) and partially inelastic (open circles,
  $\beta\!=\!0.5$) boundaries are compared. Inset: zoom in the boundary
  layer
  on the left of the dashed vertical blue line in the main panel.
 (b) Density profiles as functions of the rescaled transverse coordinate $r_\perp/R$, for increasing separation
  $R\!=\!$ 64, 128, 256, 512, and 1024, with $\beta\!=\!1$.
  Inset: Magnitude of density deviations $A_0$ (red
  symbols, see text) and rescaled zero-crossing $d_0/R$ of $\langle
  \delta \rho\rangle$ (see text) as functions of $R$. 
All the panels refer to the case with $\rho_0\!=\!2.0$, $\eta\!=\!0.21$, and $L\!=\!1024$. 
The various average profiles were obtained by a coarse-graining in boxes of unit linear size and by
averaging in time and in the longitudinal directions. 
}
\label{fig2}
\end{figure}

\emph{Coarse-grained description.---}The flocking fluid can be described
in terms of suitably coarse-grained slow hydrodynamic 
fields, i.e., the density deviations  $\delta\rho \!\equiv\! \rho
({\bf r}, t)-\rho_0$ from the mean density $\rho_0$ and the transverse velocity
fluctuations $v_\perp^2 ({\bf r}, t)$  (the longitudinal component of the velocity is
a fast mode enslaved to the slow ones \cite{TT2}).
Transverse confinement breaks
translational invariance, thus
we expect the averages 
$\langle \delta\rho \rangle$ and
$\langle v_\perp^2\rangle$ to be inhomogeneous across
the transverse direction. 
In fact, as it was already noted in Ref.~\cite{Benoit2019}, reflecting boundaries tend to suppress fluctuations in their vicinity, inducing an excess density \cite{Lowen}, such as that shown in Fig.~\ref{fig2}(a). 
For $\beta \!\!\neq\!\! 1$, these profiles turn out to coincide with those for $\beta\!\!=\!\!1$, with the sole exception of the first (non extensive) bin closest to the wall, see the inset of Fig.~\ref{fig2}(a).
One can define the extension $d_0$ of this boundary layer as the
typical distance from the wall at which density deviations change sign. 
The scaling analysis reported in
Fig.~\ref{fig2}(b) shows that, asymptotically, $d_0 \!\propto\! R$: accordingly, the
boundary layer extends well beyond the microscopic
particle-wall interaction range, being a finite fraction of 
$R$. Note, however, that the
amplitude
$A_0 \!=\! \int_0^1 dx \left | \langle \delta \rho(x, t) \rangle
\right | \!\sim\! 1/\sqrt{R}$ (where $x\!=\!r_\perp/R$) of the density
deviations decays asymptotically to zero, so that the bulk TT hydrodynamic
behavior is recovered for $R \!\to\! \infty$. 
The extensivity of $d_0$
suggests that the  long-range behavior of the mechanical pressure has a collective origin and therefore it should be accessible via the TT theory.

We now turn our attention to the mesoscopic definition of pressure in 
Eq.~\eqref{eq:pressure}, which implies
$\overline{P_b}\approx \rho_0+\langle \delta \rho \rangle$.
In order to determine
the average density at the confining walls,
we consider the TT equations in a slightly simplified form
\cite{Bartolo2017}. This retains all the essential
  non-equilibrium terms describing the advection of  
  $\rho ({\bf r}, t)$ and
  the velocity ${\bf v}({\bf r}, t)$ \cite{HydroReview}, i.e.,
\begin{align}\label{a: rho evolution}
&\partial_t \rho + \nabla \cdot (\rho{\bf v})=D_\rho \nabla^2 \rho, \\
\label{a: v evolution}
&\partial_t{\bf v}+\lambda_1
({\bf v}\cdot\nabla) {\bf v}=(\alpha - \beta {\bf v}^2) {\bf v}
-\sigma_1 \nabla \rho +D_v\nabla^2\mathbf{v}+\mathbf{f},
\end{align}
where $\mathbf{f}$ is a delta-correlated white noise of amplitude
$\Delta$.
In these equations
we have also included a nonzero density diffusion term for
$\rho$, proportional to $D_\rho$, which in the TT equations is generated by
renormalization \cite{TT3}. The (positive) coefficients $D_v$,
$\lambda_1$ and $\sigma_1$ are related, respectively, to diffusion,
advection and pressure. The Ginzburg-Landau terms (i.e., the first two on the r.h.s. of 
Eq.~\eqref{a: v evolution}, with $\alpha, \beta >0$
in the ordered phase) determine the modulus $p_0\equiv|{\bf p}_0|=\sqrt{\alpha(\rho_0)/\beta}$ of the average bulk
velocity ${\bf p}_0$.
Due to confinement, 
${\bf p}_0$ aligns with the unit vector ${\bf \hat{e}}_\parallel$ parallel to
the walls. For simplicity, we consider constant coefficients, only retaining the local density dependence of $\alpha=\alpha(\rho)$ which plays a fundamental role in the onset of the phase separated regime \cite{Bertin}. We proceed by spin-wave
  approximation \cite{Nishimori}, decomposing the velocity ${\bf v}$ into its longitudinal and transversal components, 
${\bf v} = (p_0 + \delta v_\parallel) {\bf\hat{e}}_\parallel + {\bf v}_\perp $,
and expanding in small velocity deviations from $ {\bf
  p}_0$. Taking averages 
  and projecting Eqs.~\eqref{a: rho evolution} and \eqref{a: v evolution}  
  along the transversal direction one gets
\begin{align}\label{rho1}
&\partial_\perp \langle {\bf v}_\perp \rangle \approx \rho_0^{-1}  D_\rho \partial_\perp^2 \langle \delta\rho\rangle,\\
\label{vperp}
&\frac{\lambda_1}{2} \partial_\perp \langle {\bf v}_\perp^2 \rangle =
-\sigma_1 \partial_\perp \langle \delta \rho \rangle +
D_v \partial_\perp^2 \langle {\bf v}_\perp \rangle,
\end{align}
where we have retained up to quadratic terms in the transversal velocity \cite{Note-v2} 
and used translational invariance
w.r.t.~time and the longitudinal direction, i.e., $\partial_t
\langle \cdot \rangle = \partial_\parallel \langle \cdot
\rangle=0$. 
We also discarded a higher-order contribution $\langle {\bf v}_\perp \delta
\rho\rangle$ from Eq.~\eqref{rho1}.
Note that by projecting Eq.~\eqref{a: v evolution} along 
the longitudinal direction we simply obtain
an equation which expresses the enslaving of $\langle \delta v_\parallel \rangle$ to
$\langle\delta \rho \rangle$ and $\langle {\bf v}_\perp^2\rangle$ \cite{TT3}. To zeroth order in the derivatives and to first order in $\langle \delta \rho\rangle$ it reads
\begin{equation}\label{enslave}
\langle \delta v_\parallel\rangle \approx \frac{\alpha'(\rho_0)}{2\beta p_0} \langle\delta \rho\rangle - \frac{\langle \mathbf{v}_\perp^2 \rangle}{2 p_0}\,,
\end{equation}
where $\alpha'(\rho) = d\alpha(\rho)/d\rho$.
Substituting Eq.~(\ref{rho1}) into Eq.~(\ref{vperp}) we
obtain that $\langle\delta \rho\rangle$ satisfies the equation 
\begin{equation}\label{eq:profile}
2 \rho_0^{-1}D_v D_\rho\, \partial_\perp^3 \langle\delta \rho\rangle - 2 \sigma_1 \partial_\perp \langle\delta \rho\rangle= \lambda_1 \partial_\perp \langle {\bf v}_\perp^2 \rangle.
\end{equation}
Neglecting
the nonlinear term $ \overline{w} \equiv \langle {\bf v}_\perp^2 \rangle
$ in Eq.~\eqref{eq:profile} renders the (linear) equation for the density profile derived in Ref.~\cite{Caprini2018} for confined {\it scalar} active matter, resulting in non-extensive boundary
layers. 
In the present {\it vectorial} case, instead, bulk correlations lead
to an extensive boundary layer,
due to having $\overline{w}\neq 0$ in Eq.~\eqref{eq:profile}.
Note that because of the symmetry of the transverse
confinement, the average profiles
$\overline{\delta \rho} \equiv \langle \delta\rho\rangle$ and $\overline{w}$ are even functions of $r_\bot$ w.r.t.~the mid-line of the slab at $r_\bot =R/2$. 
This implies that
$\overline{\delta \rho}(r_\perp)=\sum_{q_\perp} \widehat{\delta
    \rho}(q_\perp) \cos (2 q_\perp r_\perp)$
(and similarly for $\overline{w}(r_\perp$)), where $q_\perp=\pi n/R$,
$n$ is a positive integer,
and the Fourier transform $\widehat{\delta
    \rho}(q_\perp)$ is given by
\begin{equation}
\label{F:exp}
\widehat{\delta
    \rho}(q_\perp)=\frac{-\lambda_1\,\widehat{w}(q_\perp)}{(2
  D_v D_\rho/\rho_0)q_\perp^2+2\sigma_1}.
\end{equation}
We can derive the Fourier modes $\widehat{w}(q_\perp)$ of the nonlinear term $\overline{w}(r_\perp)$ from the 
two-point static bulk correlations in Fourier space. According to TT theory 
\cite{TT2}, one has  
\begin{equation}
\label{scaling_v} 
\langle |\hat{\bf v}_\perp ({\bf q})|^2\rangle \underset{|{\bf
    q}| \to0}{\sim} \left\{q_{\perp}^{-\psi} 
\,{\rm for}\, q_\|  \ll q_{\perp}^\xi ;\;
q_{\|}^{-\psi/\xi} \,{\rm for}\, q_\| \gg q_{\perp}^\xi  \right\},
\end{equation}
where $\psi\!\equiv\!d-1+2\chi + \xi$, while the
universal
scaling exponents $\chi\!<\!0$ and $\xi\!\leq\!1$ control, respectively, the change with length
scales of the slow field fluctuations and the spatial anisotropy between the
transverse and the longitudinal directions.

Imposing zero Dirichlet boundary conditions on the {\it instantaneous} coarse-grained field ${\bf
  v}_\perp (r_\perp,t)$ in order to model the reflecting walls and integrating out the longitudinal wave-numbers,  
we have for $q_\perp>0$ and up to an unknown constant $C>0$, 
\begin{equation}
\widehat{w}(q_\perp) = - C R^{-1} q_\perp^{-(1+2\chi)},
\end{equation}
see Ref.~\cite{SM} for details.
Substituting this expression into
Eq.~(\ref{F:exp}) and summing up over the modes of the density profile it is possible to determine $\overline{\delta \rho}(r_\perp)$ in real space.
While the resulting density profile will be discussed elsewhere
\cite{preparation1}, here we focus on its value at the boundaries $B = 0, R$.
Up to higher-order corrections in $1/R$, it turns out to be given by  (see Ref.~\cite{SM} for details),
\begin{equation}
\overline{\delta \rho}(B)=\sum_{q_\perp>0} \widehat{\delta
    \rho}(q_\perp) = \Omega\, C\left[S_\infty + \frac{\zeta (1+2\chi)}{\pi^{(1+2\chi)}}
 R^{2 \chi}\right],
 \label{eq:scal-drho}
\end{equation}
where $\Omega=\lambda_1/(2 \sigma_1)>0$, $\zeta(z)$ is the Riemann zeta function and $S_\infty>0$ is the bulk contribution to the boundary density, which determines the pressure 
in a semi-infinite system, i.e., for $R \to \infty$. 
The TT universal exponents are not known exactly \cite{TT3}, but the best large-scale numerical
estimates available in $d\!=\!2$ yield asymptotically
$\chi\!=\!-0.31(2)$ \cite{ Benoit2019}.
Inserting Eq.~\eqref{eq:scal-drho} in Eq.~\eqref{eq:pressure} yields a theoretical prediction for $\overline{P}_b(R)$ which correctly captures
the qualitative behavior of the numerical boundary pressure in
Eq.~(\ref{eq:pow}), approaching the asymptotic bulk
value from below (being $\zeta(1+2\chi)<0$).

This confirms the attractive nature of the Casimir-like force in 
Eq.~\eqref{eq:Casimir}, with an algebraic decay $\sim R^{2\chi}$, i.e., with a 
Casimir exponent $\alpha$ predicted to be $\alpha = -2\chi \approx
0.6$. 
Note, however, that the analytical argument of Ref.~\cite{TT3} would give $\alpha=2/5$ under the 
assumption that certain nonlinear bulk
contributions to TT equations are irrelevant. 
While numerics clearly shows this is not the case for large $R$,
Ref.~\cite{Benoit2019} also reports a crossover
which makes the effective value of $|\chi|$ increase upon increasing the system size,  
possibly due to the late onset of relevant nonlinear contributions. 
Overall, these considerations
are compatible with the crossover of the Casimir
exponent from $\approx 0.4$ to $\approx 0.6$ which we observed
in Fig.~\ref{fig1}(b) by excluding from the fit pressure data from smaller system sizes.

\emph{Conclusions.---}We have shown that, in the flocking phase, an active polar fluid confined
between two infinite and parallel
reflecting walls separated by a distance $R$ 
is characterized  by an excess number
density at the boundaries. This creates {\it extensive} boundary layers, as
opposed to the finite ones usually observed in confined scalar active matter
\cite{Caprini2018}. 
This behavior is ultimately induced by the
diverging bulk correlation length and by the coupling
between the slow fields of density and velocity, which characterizes
vectorial active matter. 
In turn, this leads to 
the emergence of an \emph{attractive} 
long-range force $\!\propto\! R^{2\chi}$
on the two walls. 
We based our analytical discussion on a slightly
simplified version of TT equations, but the 
same conclusion can be drawn from the complete TT theory \cite{preparation1}. 
In this latter
case too, the Casimir force turns out to be attractive
provided one considers the specific values of the transport
coefficients computed from the direct coarse-graining of microscopic
Vicsek-like models \cite{Bertin, Ihle}.
Note that our result also implies that the bulk exponent $\chi$ can be measured by the finite-size scaling of the mechanical pressure.

The non-equilibrium Casimir-like force investigated here is analogous to the critical
Casimir force $\!\propto\! T R^{-d}$ at equilibrium in $d$ spatial dimensions, which arises due to long-ranged correlated fluctuations, but with important differences. In the latter case, in fact, the algebraic dependence on $R$ is fixed by the finite-size scaling behavior of the free energy from which the force is derived (see, e.g., Ref.~\cite{KB-94}).
In the present case, instead, the long-ranged force is directly caused by a
fluctuation-induced accumulation of density at the boundaries and
turns out to be controlled by the field scaling exponent
$\chi$. 

Note finally that, being the Vicsek fluid {\it compressible}, the force discussed here bears no relation with the hydrodynamic interactions
arising in molecular fluids, due to {\it incompressibility} \cite{kundu2015}.

Here we focused our analysis on the flocking phase. 
A detailed study of the effects of confinement in other regions of the phase diagram, e.g., in the micro-phase separated band phase
\cite{Chate2,VicsekPD} is beyond the scope of this work. However, preliminary results obtained in the
disordered phase \cite{SM} indicate the emergence of a \emph{repulsive} force between the reflecting
walls with a much shorter range, possibly controlled by the
finite correlation length of fluctuations in the disordered phase. 
Correspondingly, the
density profile is depleted 
near the reflecting walls. This is at odds with what observed
in Ref.~\cite{Ray} for confined active Brownian particles (ABPs). Note, however, that
the boundaries considered there are not reflecting and do not change
the persistent directions of the ABPs.

While we considered here the active medium to be confined between parallel walls, it is both conceptually interesting and practically relevant for experiments to consider the case in which a spherical inclusion or tracer is immersed in the medium, studying how it interacts with either a flat boundary or other inclusions.
\begin{acknowledgments}
\emph{Acknowledgments.---}We warmly thank F.~Giavazzi and R.~Cerbino for having drawn
our attention on this problem and for early discussions. We
acknowledge S.~Ramaswamy and J.~Toner for
valuable discussions. GF and FG  acknowledge support from PRIN
2020PFCXPEAG. AG acknowledges
support from MIUR PRIN project ``Coarse-grained description for non-equilibrium systems and transport 
phenomena (CO-NEST)'' n.~201798CZL.
\end{acknowledgments}
\bibliography{biblio}

\begin{thebibliography}{60}
\bibitem [{not()}]{note-npara}%
  \BibitemOpen
  \href@noop {} {\bibinfo  {journal} {That is $n_\parallel \to$ sgn$(n_\parallel)\sqrt{1-\beta(1-n_\parallel^2)}$}\ }\BibitemShut {NoStop}%
\end{thebibliography}

\ifarXiv
    

\section*{Supplementary material}

# Supplementary materials to the manuscript
# "Strong Casimir-like Forces in Flocking Active Matter"

Giuseppe Fava,[1,2] Andrea Gambassi,[3] and Francesco Ginelli[1,2]

[1]*Dipartimento di Scienza e Alta Tecnologia and Center for Nonlinear and Complex Systems, Università degli Studi dell'Insubria, Como, Italy*
[2]*INFN sezione di Milano, Milano, Italy*
[3]*SISSA — International School for Advanced Studies and INFN, via Bonomea 265, 34136 Trieste, Italy*

## S.I. ANALYTICAL DETERMINATION OF THE PRESSURE ON THE BOUNDARIES

In this section we provide the details of the analysis in Fourier space which allows one to derive the density deviation $\overline{\delta\rho}(B)$ at the boundaries $B$ described by Eq. (14) of the main text from the Fourier modes $\widehat{\delta\rho}(q_\perp)$ of the density profile $\langle\delta\rho\rangle$ given by Eq. (11) therein.

### A. Boundary conditions and slow fields expansion

In order to model reflecting walls, we adopt zero Dirichlet boundary conditions for the instantaneous transversal velocity field $v_\perp(\mathbf{r}, t)$. In the limit of infinite longitudinal extension $L \to \infty$, the field $v_\perp(\mathbf{r}, t)$ can be expressed in terms of its Fourier transform $\hat{v}_\perp(\mathbf{q}, t)$ as

$$v_\perp(\mathbf{r}, t) = \int_{-\infty}^{\infty} dq_\parallel \sum_{q_\perp > 0} \hat{v}_\perp(\mathbf{q}, t) \sin(q_\perp r_\perp) e^{iq_\parallel r_\parallel}, \quad \text{(S.1)}$$

where $\mathbf{q} = (q_\parallel, q_\perp)$, with $q_\parallel \in \mathbb{R}$ while $q_\perp$ takes discrete values $q_\perp = q_\perp(n) = \pi n/R$, with $n$ a positive integer. The Fourier modes are given by

$$\hat{v}_\perp(\mathbf{q}, t) = \frac{1}{\pi R} \int_{-\infty}^{\infty} dr_\parallel \int_0^R dr_\perp v_\perp(\mathbf{r}, t) \sin(q_\perp r_\perp) e^{-iq_\parallel r_\parallel}. \quad \text{(S.2)}$$

An analogous expansion of the delta-correlated noise term $\mathbf{f}(\mathbf{r}, t)$ appearing in the TT equations (see Eq. (6) of the main text) leads to Fourier space correlations

$$\langle \hat{\mathbf{f}}_i(\mathbf{q}, t) \hat{\mathbf{f}}_j(\mathbf{q}', t') \rangle = \frac{\Delta}{\pi R} \delta_{q_\perp, q'_\perp} \delta(q_\parallel - q'_\parallel) \delta(t - t') \delta_{i,j}, \quad \text{(S.3)}$$

which, in turn, constrain Fourier space correlations of the slow fields [3].

In particular, for equal times,

$$\langle \hat{v}_\perp(\mathbf{q}) \hat{v}_\perp(\mathbf{q}') \rangle = \frac{1}{\pi R} \langle |\hat{v}_\perp(\mathbf{q})|^2 \rangle \delta_{q_\perp, q'_\perp} \delta(q_\parallel - q'_\parallel). \quad \text{(S.4)}$$

The behavior of these correlations of the transversal velocity for small $|\mathbf{q}|$ is determined by the bulk TT theory in the flocking phase[1], which predicts

$$\langle |\hat{v}_\perp(\mathbf{q})|^2 \rangle \sim \left\{ q_\perp^{-\psi} \text{ for } q_\parallel \ll q_\perp^\xi;\ q_\parallel^{-\psi/\xi} \text{ for } q_\parallel \gg q_\perp^\xi \right\}, \quad \text{(S.5)}$$

where the exponents $\psi$ and $\xi$ are defined after Eq. (12) in the main text. In Eq. (S.5) the subscripts $\parallel$ and $\perp$ refer to the directions which are, respectively, parallel and perpendicular to the polarization $\mathbf{p}_0$ of the flock in the bulk. We recall that in the confined system, $\mathbf{p}_0$ is naturally parallel to the confining walls.

We now consider the average fluctuations of the transversal velocity. Due to translational invariance along the longitudinal direction, from Eqs. (S.1) and (S.4), one finds

$$\langle v_\perp^2(r_\perp) \rangle = \frac{1}{\pi R} \sum_{q_\perp > 0} \sin^2(q_\perp r_\perp) \int dq_\parallel \langle |\hat{v}_\perp(\mathbf{q})|^2 \rangle. \quad \text{(S.6)}$$

Integration over the longitudinal modes $q_\parallel$ gives

$$\int_{-\infty}^{\infty} dq_\parallel \langle |\hat{v}_\perp(\mathbf{q})|^2 \rangle \approx c_1 \int_0^{q_\perp^\xi} dq_\parallel q_\perp^{-\psi} + c_2 \int_{q_\perp^\xi}^{\infty} dq_\parallel q_\parallel^{-\psi/\xi} \quad \text{(S.7)}$$
$$= 2\pi C q_\perp^{-\gamma},$$

where we used Eq. (S.5) and we introduced $\gamma \equiv \psi - \xi = d - 1 + 2\chi$. Numerical estimates [1] and analytical predictions [2] in $d = 2$ give $0 > \chi > -1/2$ and therefore $0 < \gamma < 1$. Note also that the two integrals on the r.h.s. of the first line of Eq. (S.7) are weighted by possibly different unknown multiplicative factors $c_1$ and $c_2$, which result in an inconsequential positive constant $2\pi C$, the exact value of which is irrelevant in the following. Substituting this expression into Eq. (S.6), we finally obtain

$$\langle v_\perp^2(r_\perp) \rangle = \frac{C}{R} \sum_{q_\perp > 0} q_\perp^{-\gamma} [1 - \cos(2q_\perp r_\perp)]. \quad \text{(S.8)}$$

From Eq. (S.8) we readily identify the coefficients of the Fourier series of $\langle v_\perp^2(r_\perp) \rangle$, taking into account that

---

[1] Repeating the analysis done in Ref [2], it is easy to show that the simplified form of the TT equations introduced in the main text have the same linearized structure and nonlinearities of the complete TT equations. Accordingly, their correlations functions are characterized by the same scaling behavior.

because of symmetry, it is an even function for spatial reflection about the mid-line of the slab, located at $r_\perp = R/2$ (see also the main text):

$$\langle v_\perp^2(r_\perp)\rangle = \hat{w}_0 + \sum_{q_\perp > 0} \hat{w}(q_\perp) \cos(2q_\perp r_\perp). \qquad (S.9)$$

Accordingly, the coefficients are given by

$$\hat{w}_0 = \frac{1}{R}\int_0^R dr_\perp \langle v_\perp^2(r_\perp)\rangle = \frac{C}{R}\sum_{q_\perp>0} q_\perp^{-\gamma}, \qquad (S.10)$$

and

$$\hat{w}(q_\perp) = \frac{2}{R}\int_0^R dr_\perp \langle v_\perp^2(r_\perp)\rangle \cos(2q_\perp r_\perp) = -\frac{C}{R} q_\perp^{-\gamma}, \qquad (S.11)$$

for $q_\perp > 0$, which coincides with Eq. (13) of the main text. Note that the formal divergence of $\hat{w}_0$ is actually regularized by a microscopic short wavelength cutoff, implicitly present in Toner-Tu theory [3, 4].

The expression of $\hat{w}(q_\perp)$ in Eq. (S.11) can now be used in order to calculate – summing up the Fourier modes of the density profile given by Eq. (11) of the main text – the value $\overline{\delta\rho}(B)$ acquired at the boundary by the average density $\langle\delta\rho(r_\perp)\rangle$. By fixing $r_\perp = 0$ or, equivalently, $r_\perp = R$ one has (due to the particle number conservation, the zero mode of the density vanishes)

$$\langle\delta\rho(0)\rangle = \frac{C}{R}\sum_{q_\perp>0} \frac{\Omega}{1+\Gamma^2 q_\perp^2}\frac{1}{q_\perp^\gamma} \equiv C\,\Omega\,S_R, \qquad (S.12)$$

where we defined

$$\Omega = \frac{\lambda_1}{2\sigma_1}, \quad \Gamma = \sqrt{\frac{D_v D_\rho}{\sigma_1 \rho_0}}, \qquad (S.13)$$

and introduced the sum

$$S_R \equiv \frac{1}{R}\sum_{n=1}^\infty \frac{1}{1+\Gamma^2 q_\perp^2(n)}\frac{1}{q_\perp^\gamma(n)}, \qquad (S.14)$$

where $q_\perp(n)$ is given after Eq. (S.1). The dependence of $S_R$ on the (large) distance $R$ can be evaluated as explained in Sec. S.I B and turns out to be

$$S_R = S_\infty + \frac{\zeta(\gamma)}{\pi^\gamma}\frac{1}{R^{1-\gamma}} + \mathcal{O}\left(\frac{1}{R^{3-\gamma}}\right), \qquad (S.15)$$

where $\zeta(\gamma)$ is the Riemann zeta function. Note that $0 < \gamma < 1$ in the case we are interested in here. Accordingly, $\zeta(\gamma) < 0$ and therefore $S_R$ approaches $S_\infty$ from below. In the previous expression

$$S_\infty = \left[2\Gamma^{1-\gamma}\cos\left(\frac{\pi\gamma}{2}\right)\right]^{-1}, \qquad (S.16)$$

corresponds to the contribution of the integral which emerges in the limit $R \to \infty$ from interpreting $S_R$ as a Riemann sum. It determines the *bulk* pressure acting on the confining walls.

Substituting Eq. (S.15) into Eq. (S.12) leads immediately to Eq. (14) of the main text.

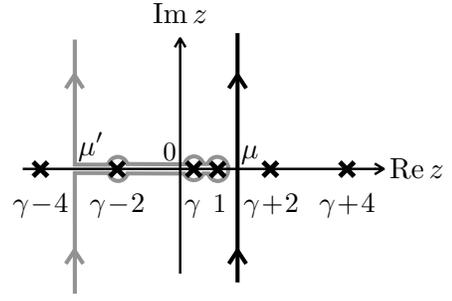

FIG. S.1: Contours in the complex plane $z \in \mathbb{C}$ along which the integral in Eq. (S.21) is calculated for determining $S_R$. The black vertical contour with $\text{Re}\,z = \mu$ is the one indicated in Eq. (S.21), which can be deformed continuously into the equivalent grey contour used in order to determine the expansion of $S_R$ upon increasing $R$ in Eq. (S.22). The crosses on the real axis indicate the poles of the integrand, which is otherwise an analytic function.

### B. Calculation of $S_R$

The sum $S_R$ in Eq. (S.14) can be conveniently evaluated by adapting to the present case the general strategy discussed, e.g., in Ref. [5]. In particular, we first rewrite $S_R$ as

$$S_R = \frac{1}{R}\left(\frac{R}{\pi}\right)^\gamma \sum_{n=1}^\infty f(n), \qquad (S.17)$$

in which we take into account the definition of $q_\perp(n)$ given after Eq. (S.1) and we define

$$f(n) = \frac{1}{n^\gamma}\frac{1}{1+(\Gamma\pi n/R)^2}. \qquad (S.18)$$

Then, we introduce the Mellin transform $\hat{f}(z)$ of $f(n)$ in the complex plane $z \in \mathbb{C}$, which is given by

$$\hat{f}(z) = \int_0^\infty dx\, f(x) x^{z-1} = \frac{\pi}{2}\frac{(\Gamma\pi/R)^{\gamma-z}}{\sin(\pi(z-\gamma)/2)}, \qquad (S.19)$$

with $\gamma < \text{Re}\,z < 2+\gamma$. This condition on $\text{Re}\,z$ guarantees that the integral above is convergent and therefore that $\hat{f}(z)$ is an analytic function within that strip in the complex plane. The function $f(n)$ in Eq. (S.17) can now be expressed in terms of $\hat{f}(s)$ as

$$f(n) = \int_{\mu-i\infty}^{\mu+i\infty}\frac{dz}{2\pi i}\,\hat{f}(s)n^{-z}, \qquad (S.20)$$

where $\mu$ is chosen within the domain of analiticity of $\hat{f}(z)$, i.e., $\gamma < \mu < 2+\gamma$. After this substitution, the remaining sum over the positive integer $n$ renders the Riemann function $\zeta(z)$ provided that this sum converges, which additionally requires $\mu = \text{Re}\,z > 1$. Accordingly, one can eventually write

$$S_R = \frac{1}{R}\left(\frac{R}{\pi}\right)^\gamma \int_{\mu-i\infty}^{\mu+i\infty}\frac{dz}{2\pi i}\,\hat{f}(z)\zeta(z), \qquad (S.21)$$

where $1 < \mu < 2 + \gamma$. Note that the integration contour above consists of a vertical line in the complex plane with fixed $\operatorname{Re} z = \mu$, which is represented by the black path in Fig. S.1. Noting that $\hat{f}(z) \propto R^{z-\gamma}$, the expansion of $S_R$ in decreasing powers of $R$ — suitable for considering the case of large $R$ we are interested in — can be obtained by shifting the integration contour in Eq. (S.21) towards smaller values of $z$, i.e., towards the left. In doing so, one has to account for the poles that the integrand has on the real axis. They are given by: (i) the pole of $\zeta(z)$ located at $z = 1$, with residue $\mathcal{R}^* = 1$; (ii) the poles of $\hat{f}(z)$ located at $\gamma - 2m$, with $m \in \mathbb{Z}$ and residues $\mathcal{R}_m = (-1)^m (\Gamma \pi/R)^{2m}$. These poles are indicated by the crosses on the real axis in Fig. S.1 and they are actually the only singularities of the integrand, which is otherwise an analytic function of $z$. Accordingly, the integral in Eq. (S.21) can be equivalently calculated, for example, along the grey contour depicted in Fig. S.1, in which the contributions of the horizontal portions of the path actually cancel each other in pairs. As a result, the integral along the grey contour is the same as that in Eq. (S.21) but with $\mu$ replaced by $\mu'$, plus the contributions of the portions of the paths encircling the poles, which — due the orientation of the path and to Cauchy residue theorem — render the residues at those poles. Accordingly, one finds

$$S_R = \frac{1}{R}\left(\frac{R}{\pi}\right)^\gamma \left[\hat{f}(z=1)\mathcal{R}^* + \zeta(\gamma)\mathcal{R}_0 + \mathcal{O}(\mathcal{R}_1)\right], \quad (S.22)$$

where $\hat{f}(z=1) \propto R^{1-\gamma}$, $\mathcal{R}_0 = \mathcal{R}^* = 1$, and $\mathcal{R}_1 \sim R^{-2}$. Taking into account Eq. (S.19), this expansion yields Eq. (S.15), upon identifying the first term on the r.h.s. with the $R$-independent constant $S_\infty$ in Eq. (S.16) (see also Eqs. (S.18) and (S.19)) and the second term with the Casimir-like, leading finite-size contribution.

## S.II. ADDITIONAL NUMERICAL SIMULATIONS

In this section we provide additional numerical evidence of the robustness and, to a certain extent, of the universality of the results presented in the main text.

### A. Universality of the pressure scaling

*Alternative definition of the mechanical pressure* — As we argued in the main text, the exact expression of the exchanged momentum at the reflecting boundaries depends on the details of the collisions between the particles and the confining wall. In the main text we considered the pressure $\overline{P}_b(R)$ defined only on the basis of the number of collisions per unit length and unit time.

Here we consider, instead, an alternative definition of the microscopic mechanical pressure $\overline{P}_b^{\mathrm{MOM}}(R)$ as the average momentum $\Delta p$ exchanged per unit length and

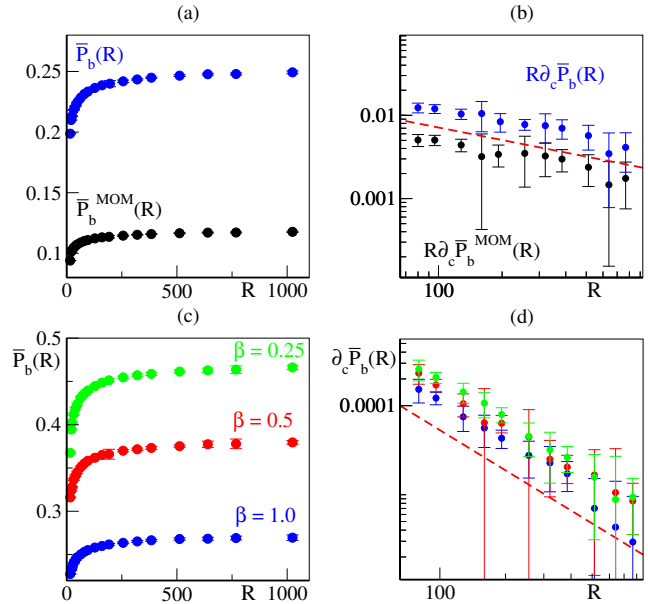

FIG. S.2: (a) Numerical estimate of $\overline{P}_b^{\mathrm{MOM}}(R)$ (black dots) and $\overline{P}_b(R)$ (blue) for fully reflecting boundaries, i.e., $\beta = 1$, with $\eta = 0.21$, $\rho_0 = 2.0$, $v_0 = 0.5$, and $L = 1024$. (b) Rescaled centered finite differences of the pressures presented in panel (a) as a function of $R$ compared to an algebraic decay $\propto R^{-\alpha}$ with exponent $\alpha = 0.5$ (red dashed line). (c) Numerical estimate of $\overline{P}_b(R)$ for various values of the reflection parameter $\beta = 1.0$ (blue), 0.5 (red), 0.25 (green). The remaining parameters are $\eta = 0.24$, $v_0 = 0.5$, and $L = 1024$. (d) Corresponding centered finite differences compared to the algebraic decay $\sim R^{-\alpha-1}$ with $\alpha = 0.5$ (red dashed line).

unit time. We assume that $\Delta p$ is entirely exchanged with the boundary (and not with the surrounding viscous fluid/substrate), so that $\Delta p = 2\beta v_0 n_\perp$ whenever the particle collides with one of the boundaries. Here $\beta$ is the reflection parameter defined in the main text. If, at a certain time $t$, there are $M(t)$ such collisions with momentum exchange $\Delta p_n$, for $n = 1, \ldots, M(t)$, the instantaneous mechanical pressure, considering both walls, can be defined as

$$P_b^{\mathrm{MOM}}(t) \equiv \frac{1}{2L} \sum_{n=1}^{M(t)} |\Delta p_n|. \quad (S.23)$$

We then take a long-time average of this pressure, yielding $\overline{P}_b^{MOM} = \langle P_b^{MOM}(t) \rangle_t$. Figure S.2(a)-(b) shows that $\overline{P}_b^{\mathrm{MOM}}(R)$ is actually characterized by the same scaling with the wall distance $R$ as the one exhibited by $\overline{P}_b(R)$ defined in Eq. (2) of the main text.

Indeed, expressing the boundary pressure $\overline{P}_b$ as a function of the relevant hydrodynamics fields $\rho$, and $\mathbf{v}$ and expanding at the first order around $\rho_0$ and $\mathbf{p}_0$ [6] yields,

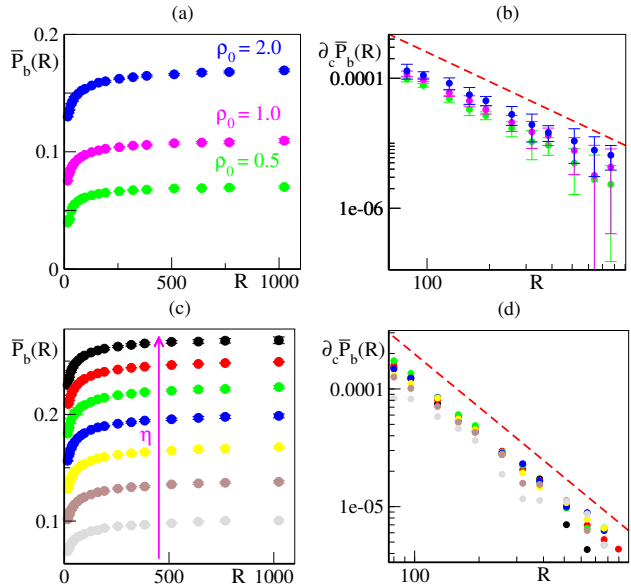

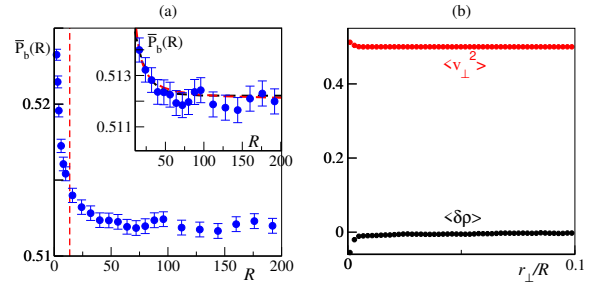

FIG. S.4: Pressure and fields behavior in the confined disordered phase, with parameters $\beta = 1$, $\eta = 0.60$, $\rho_0 = 2$, $v_0 = 0.5$, and $L = 512$. (a) Mechanical pressure $\overline{P}_b(R)$ as a function of the transverse separation $R$. The vertical dashed red line corresponds to the correlation length $\xi$ in the disordered phase (see text). Inset: exponential (black dashed line, $\overline{P}_b(R) \approx 0.5 + 0.005\,e^{-0.07R}$) and algebraic ($\overline{P}_b(R) \approx 0.5 + 0.06\,R^{-1.3}$, red dashed line) fits of the function $\overline{P}_b(R)$ for $R$ larger than the correlation length $\xi$ in the disordered phase. (b) Average density deviations (black) and transverse velocity fluctuations (red) as functions of the rescaled transverse coordinate $r_\perp/R$ (zoom of the near-wall behavior).

FIG. S.3: (a) Numerical estimate of $\overline{P}_b(R)$ for various values of the density $\rho_0 = 2.0$ (blue), $\rho_0 = 1.0$ (magenta) and $\rho_0 = 0.5$ (green). The remaining parameters are $\beta = 1.0$, $\eta = 0.12$, $v_0 = 0.5$, and $L = 1024$. (b) Corresponding centered finite differences compared to the algebraic decay $\propto R^{-\alpha-1}$ with $\alpha = 0.5$ (red dashed line). (c) Time-averaged mechanical pressure exerted by the polar fluid on the confining walls as a function of their separation $R$ for $\eta \in [0.06, 0.09, 0.12, 0.15, 0.18, 0.21,$ and $0.24]$. (d) Centered finite difference of $\overline{P}_b(R)$ as a function of $R$. The dashed red line mark corresponds to an algebraic decay $\propto R^{-\alpha-1}$ with $\alpha = 0.5$. Error bars are omitted for the sake of clarity.

after averaging,

$$\overline{P}_b(\rho, \mathbf{v}) \approx P_b(\rho_0, p_0) + \quad\quad\quad (S.24)$$
$$+\ \partial_\rho P_b(\rho_0, p_0)\langle\delta\rho\rangle + \partial_{v_\perp} P_b(\rho_0, p_0)\langle v_\perp\rangle\,.$$

The third term on the r.h.s vanishes at the boundary due to the presence of reflecting walls, which we model by zero Dirichlet boundary condition for the transversal velocity $v_\perp$. The resulting expression for the pressure depends only on the density deviations and thus should exhibit the same scaling as the average mechanical pressure given by Eq. (2) of the main text.

*Partially reflecting boundaries* — We also consider the case in which the active particles have partially reflecting collisions with the boundaries.

Simulations with $\beta = 0.5$ and $0.25$, reported in Fig. S.2(c)-(d), show that the exponent $\alpha$ of the algebraic approach $\propto R^{-\alpha}$ of the mechanical pressure $\overline{P}_b(R)$ to the bulk value $\overline{P}_b(R \to \infty)$ upon increasing $R$ is not affected by the value of $\beta$. This confirms a certain degree of universality for repelling confinements, with post-collision particles moving away from the wall.

*Exploration of the flocking phase* — Here we investigate the dependence of the mechanical pressure $\overline{P}_b(R)$ on the two control parameters: the global density $\rho_0$ and the amplitude of the microscopic noise $\eta$.

In Fig. S.3(a) we show the mechanical pressure $\overline{P}_b(R)$ as obtained for three different values of the particles density, i.e., for $\rho_0 = 2.0, 1.0$, and $0.5$, keeping the value of the microscopic noise $\eta$ fixed. We find that the bulk pressure $\overline{P}_b(R \to \infty)$ increases monotonously with the global density $\rho_0$, while the value of the *Casimir* exponent $\alpha$ does not depend on $\rho_0$ (see Fig. S.3(b)), as long as the fluid is within the flocking phase [7].

Furthermore, we also verified that the dependence of the mechanical pressure $\overline{P}_b(R)$ on the value of the microscopic noise $\eta$, now keeping the value of the global density $\rho_0$ fixed.

The results of these simulations are reported in Fig. S.3(c)-(d). The bulk pressure $\overline{P}_b(R \to \infty)$ is found to increase monotonously also with the amplitude of the noise $\eta$ while no significant deviation from the *Casimir* exponent $\alpha = 0.5$ is observed (see Fig.. S.3(d)).

Overall, this confirms the validity of our result for a quite broad set of parameters within the polar liquid phase.

### B. Pressure in the disordered phase

The long-range nature of the Casimir-like force between the two (partially) reflecting confining walls is connected to the long-range correlations which characterize the ordered phase [3]. As a counterexample, we have also performed numerical simulations in the disordered gas-like phase where, in the absence of spontaneous sym-

metry breaking, correlations are short-range.

In particular, we have considered the case $\rho_0 = 2.0$ and noise strength $\eta = 0.60$. These values ensure that the system is in the disordered gas phase [7], avoiding the region of parameter in which the system is micro-phase separated, which we do not discuss here. The behavior of the confined disordered system, presented in Fig. S.4, is qualitatively different from that obtained in the ordered phase discussed above and in the main text. In particular, Fig. S.4(a) features a change of the sign of the finite-size contribution to the total boundary pressure, resulting in a net *repulsive* force between the two boundaries. Moreover, this force has a significantly shorter spatial range, as shown by the fast decay of the pressure to its asymptotic value upon increasing $R$.

As we have noted, the disordered phase is characterized by a *finite* correlation length $\xi$, which we have estimated numerically from the equal-time correlations of the transverse velocity and which is indicated in Fig. S.4(a) as a dashed red vertical line. Once the transverse separation $R$ exceeds this finite bulk correlation length $\xi$, the decay of the mechanical pressure to its bulk value $\overline{P}_b(\infty)$ becomes compatible with a fast exponential decay.

Finally, we also note that the dependence of the density and of the transverse velocity fields on the distance from a wall, shown in Figure S.4(b), is much less pronounced than in the flocking phase, with a comparatively small boundary layer, characterized by a moderate decrease of the density and increase of velocity fluctuations near the boundaries. This behavior is the opposite of what is observed in the flocking phase; at the moment we cannot offer a theoretical explanation of these preliminary numerical results.


\fi
  
\end{document}